\documentclass[12pt]{article}
\usepackage{amssymb,amsmath}
\usepackage{graphicx}
\usepackage{color}

\newcommand{\diag}{\mathrm{diag}}
\newcommand{\Ncal}{\mathcal{N}}

\newcommand{\nbox}{{\,\lower0.9pt\vbox{\hrule \hbox{\vrule height 0.2 cm \hskip 0.19 cm \vrule height 0.2 cm}\hrule}\,}}

\def\href#1#2{#2}

\begin{document}
\begin{titlepage}
\hfill
\vbox{
    \halign{#\hfil         \cr
           hep-th/0611019  \cr
           } 
      }  
\vspace*{20mm}
\begin{center}
{\Large \bf Little String Theory from Double-Scaling Limits of Field Theories}

\vspace*{15mm}
\vspace*{1mm}

{Henry Ling, Hsien-Hang Shieh, Greg van Anders}

\vspace*{1cm}

{Department of Physics and Astronomy, University of British Columbia\\
	6224 Agricultural Road, Vancouver, B.C., V6T 1W9, Canada}

\vspace*{1cm}
\end{center}

\begin{abstract}
We show that little string theory on $S^5$ can be obtained as double-scaling
limits of the maximally supersymmetric Yang-Mills theories on $R\times S^2$ and
$R\times S^3/Z_k$.  By matching the gauge theory parameters with those in the dual
supergravity solutions found by Lin and Maldacena, we determine the limits in the gauge
theories that correspond to decoupling of NS5-brane degrees of freedom. We find that for the theory on $R\times S^2$, the 't Hooft
coupling must be scaled like $\ln^3 N$, and on $R\times S^3/Z_k$, like $\ln^2N$.
Accordingly, taking these limits in these field theories gives Lagrangian
definitions of little string theory on $S^5$.
\end{abstract}

\end{titlepage}

\vskip 1cm
\section{Introduction}
Type IIA little string theory \cite{lst1} describes the decoupling limit of
NS5-branes in type IIA string theory in the limit where $g_s$ is taken to zero
at fixed $\alpha'$. The remaining degrees of freedom are believed to be
described by a non-gravitational six-dimensional theory. The infrared limit of
this theory is known to be the $(0,2)$ conformal field theory, but in general
the theory is non-local (see \cite{lst} for a review).

Little string theory has a DLCQ formulation \cite{dlcq} as well as a
deconstruction description \cite{dcon}, however it has mainly been analyzed
through its gravity dual. This gravity dual is the near-horizon limit of the
NS5-brane solution of type IIA string theory. For large $r$, where the IIA
picture is valid, this is given by
\begin{equation}
\begin{split}
ds^2 &= N_5 \alpha'(-dt^2 + d\vec{x}_5^2 + dr^2 + d \Omega_3^2) \\
e^{\phi} &= g_s e^{-r},
\end{split}
\end{equation}
with $N_5$ units of $H$ flux through the $S^3$. This description is also
difficult to work with, however, since the linear dilaton sends the theory to
strong coupling in the infrared region of the geometry.

Recently, Lin and Maldacena \cite{lm} found a supergravity solution in
which the flat five-dimensional part of the geometry along the worldvolume of
the NS5-branes is replaced with an $S^5$. For large radius this takes the form
\begin{equation}
\begin{split}
ds^{2} &= N_5 \alpha'[2r(-dt^2+d\Omega_5^2)+dr^2+d\Omega_3^2] \\
e^\Phi &= g_s e^{-r} .
\end{split}
\end{equation}
This solution contains a linear dilaton and an $S^3$ with $N_5$ units of
$H$-flux, implying we can think of this as describing NS5-branes on $S^5$.
Interestingly, this supergravity solution has some features that make it more
tractable than the solution corresponding to flat NS5-branes. The maximum
values of the dilaton and the curvature are both tunable so that the
supergravity description is valid everywhere.

This solution is one of a family of solutions of type IIA supergravity
preserving $SU(2|4)$ symmetry constructed by Lin and Maldacena \cite{lm}.
Each supergravity solution is constructed from the electrostatic potential of an
axisymmetric arrangement of charged conducting disks in three
spatial dimensions. The above NS5-brane solution is obtained from the
electrostatic potential between two infinitely large disks.

\begin{figure} \label{3C}
\begin{center}
\includegraphics{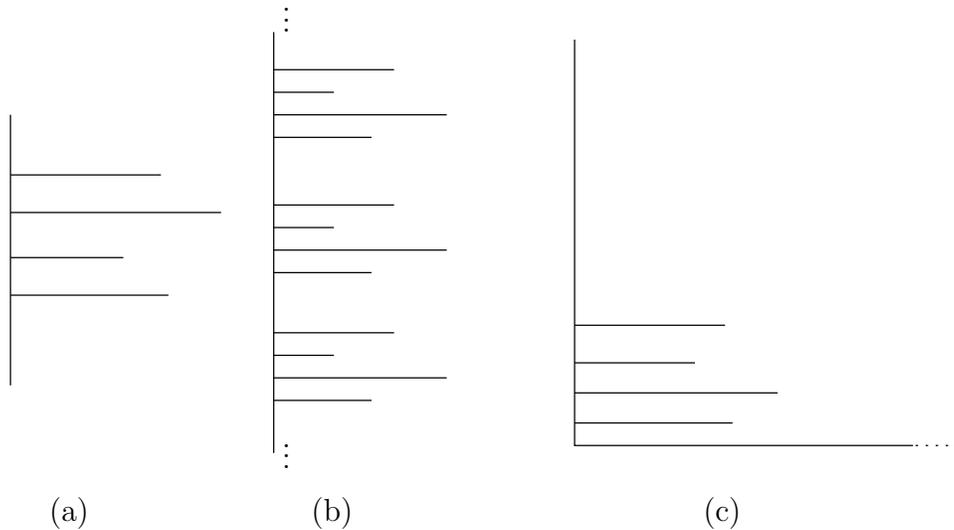}
\end{center}
\caption{The three generic types of electrostatics configurations. The isolated
set of disks in (a) is a configuration dual to a vacuum of SYM theory on $R\times S^2$ with sixteen
supercharges. The periodic configuration in (b) is dual to a vacuum of $\Ncal=4$ SYM theory
on $R\times S^3/Z_k$. The set of disks above an infinite conducting plane in (c) is
dual to a vacuum of the PWMM.}
\end{figure}

According to the proposal in \cite{lm}, the supergravity solutions arising from
configurations with a finite number of disks correspond to the (classically
degenerate) vacua of super-Yang-Mills theory on $R\times S^2$ with sixteen
supercharges. Configurations with an infinite number of disks, arranged in a
periodic fashion, correspond to the vacua of $\mathcal{N}=4$ super-Yang-Mills on
$R\times S^3/Z_k$. Finally, configurations with one infinitely large disk, and
a finite number of disks above it, correspond to the vacua of the plane wave
matrix model (see figure \ref{3C}). The relations among these field theories
have been discussed in \cite{msv,itt,istt}.

The supergravity picture suggests an interesting connection between these three
gauge theories with $SU(2|4)$ symmetry and little string theory. Lin and
Maldacena showed that the supergravity solutions dual to the various vacua of
these field theories generally contain throats with non-contractible $S^3$s
permeated by $H$-flux, which can be associated with NS5-brane degrees of
freedom. In the limit that the throats containing the non-contractible $S^3$s
with $H$-flux become infinitely large, the NS5-brane degrees of freedom will
decouple. The remaining geometry should be the above Lin-Maldacena NS5-brane
solution dual to little string theory on $S^5$ \cite{lm}. In the language of
the dual theories, this suggests that little string theory may be obtained from
suitable limits of the three gauge theories.

In \cite{lmsvv} the supergravity dual of a simple vacuum of the plane-wave model
was considered, and the required limit that gives the Lin-Maldacena solution was
explicitly determined. By matching the parameters of the plane wave matrix model
with those of the electrostatics configuration, it was proposed that little
string theory on $S^5$ may be obtained from a double-scaling limit of the plane
wave matrix model.

In this paper, we extend the work of \cite{lmsvv} and perform a similar analysis
for the SYM theory on $R\times S^2$ and $\mathcal{N}=4$ SYM theory on
$R\times S^3/Z_k$. We solve the electrostatics problems corresponding to
specific simple vacua of these field theories and determine the scaling of
parameters in the supergravity solutions that is required to obtain the
Lin-Maldacena solution for NS5-branes on $S^5$. By considering the matching
between the parameters in the field theories and those in the corresponding
electrostatics problems, we thereby determine the precise scaling of the gauge
theory parameters that is required to obtain little string theory on $S^5$.
The proposed prescriptions are found to be double-scaling limits, similar to the
one found in the case of the plane wave matrix model \cite{lmsvv}. Whereas in
the plane wave matrix model case it was found that the 't Hooft coupling must be
scaled like $\ln^4 N$ \cite{lmsvv}, we will show below that for the SYM theories
on $R\times S^2$ and $R\times S^3/Z_k$ the 't Hooft coupling must be scaled
like $\ln^3 N$ and $\ln^2 N$ respectively.

\section{The gauge theories and their dual supergravity solutions}
 In \cite{lm}, Lin and Maldacena found a class of solutions of type IIA
supergravity with $SU(2|4)$ symmetry depending on one single function $V$.
This function $V$ solves the three dimensional Laplace equation and satisfies
the same boundary conditions as the electrostatic potential of an axisymmetric arrangement of charged conducting disks in a background electric
field. By specifying the positions and sizes of the conducting disks, the
charges on the disks, and the asymptotic form of $V$ at infinity, $V$ is
determined uniquely. Each different specification of these parameters leads to
a different $V$, however not all such choices give rise to physically acceptable
supergravity solutions. Flux quantization in the supergravity solution tells us
that the charges on the disks and the spacing between disks are quantized.
Positive-definiteness of various metric components in the supergravity solutions
imposes constraints on the form of the asymptotic potential. Finally, the
regularity of the supergravity solutions tells us that the surface charge
density on the disks must vanish at the edge of the disks. This final condition
suggests that for a fixed asymptotic potential, the positions, charges, and
sizes of the disks cannot be independently specified. For example, the sizes of
the disks may be fixed once the other parameters are freely specified. For an
extensive discussion of the general properties of these supergravity solutions
see \cite{lm}.

Here we are interested in the supergravity solutions dual to the vacua of the
SYM theory on $R\times S^2$ and $\mathcal{N}=4$ SYM theory on $R\times S^3/Z_k$.
For all vacua of these two field theories, Lin and Maldacena determined the
asymptotic form of $V$ to be $W_0(r^2-2z^2)$, where $W_0>0$. The choice of
vacuum is then given by specifying the charges, positions, and sizes of the
disks. Let us review in some detail the connection between these parameters for
the supergravity solutions and the parameters defining the field theory vacua.

First consider $\mathcal{N}=4$ SYM on $R\times S^3/Z_k$. The space $S^3/Z_k$
can be described most directly by choosing coordinates on the unit $S^3$ such that
the metric takes the form
\begin{equation}
d\Omega_3^2 = \frac{1}{4}\left[ (2 d\psi +\cos\theta d\phi)^2 + d\theta^2 + \sin^2\theta d\phi^2 \right],
\end{equation} 
where the $\psi$ coordinate is $2\pi$ periodic, and $\theta$, $\phi$ are the
usual coordinates for $S^2$. Then the orbifold $S^3/Z_k$ is obtained by
identifying $\psi \sim \psi + 2\pi/k$. The vacua of this field theory are given
by the space of flat connections on $S^3/Z_k$. Up to gauge transformations,
these are of the form $A = -\diag(n_1, n_2, \dots, n_N)\ d\psi$, where
$e^{2\pi n_i/k}$ are $k$-th roots of unity (clearly, to label the vacua
uniquely, we should restrict the values of the integers $n_i$ to be in some
fixed interval of length $k$). To understand intuitively how these vacua map to
configurations of disks in the electrostatics problem, consider the field
theory as a theory of D3-branes wrapped on an $S^3/Z_k$. Now apply a T-duality
transformation in the isometry direction $\psi$. The T-dual coordinate
$\tilde{\psi}$ is periodic $\tilde{\psi} \sim \tilde{\psi} + 2\pi k$, and the
background gauge field is mapped to an arrangement of D2-branes located at the
positions $\tilde{\psi}=2\pi n_1, 2\pi n_2,\dots, 2\pi n_N$ (along with their
images under translations by integer multiplies of $2\pi k$). Naturally, this
suggests that the dual supergravity solution is obtained by considering a
periodic configuration of disks with period proportional to $k$. The integers
$n_i$ that specify the gauge theory vacuum now determine the positions and
charges of the disks within one period in the obvious manner. Presumably, the
sizes of the disks are then fixed by demanding regularity of the supergravity
solution. In rest of this paper, we will be interested in the simplest vacuum
state of the theory, given by the trivial gauge field $n_i=0$ for all $i$. In
the normalization conventions of Lin and Maldacena \cite{lm}, the dual
supergravity solution is generated by the axisymmetric electrostatic
potential $V(r,z)$ for an arrangement of equal-sized disks at $z=(\pi/2)km$ for
all integers $m$, where the charge on each disk is $Q=(\pi^2/8)N$.

Now we consider the case of the SYM theory on $R\times S^2$. As discussed in
\cite{lm}, we can think of this theory as $\mathcal{N}=4$ SYM on
$R\times S^3/Z_k$, in the limit where $k \rightarrow \infty$ and
$g^2_{YM3}\rightarrow 0$ while keeping $g^2_{YM3} k$
fixed. Up to a numerical constant, the limiting value of $g^2_{YM3}k$ is the coupling
$g^2_{YM2}$.  If we start with a vacuum in
the $S^3/Z_k$ theory with background gauge field
$A = -\diag(n_1, n_2, \dots, n_N)d\psi$ and take $k\rightarrow \infty$ with the
integers $n_i$ fixed, then we obtain a vacuum of the $S^2$ theory with a vacuum
expectation value for one of the adjoint scalars
$\Phi = -\diag(n_1, n_2, \dots, n_N)$ and a background gauge field with
associated flux $F=dA=\Phi \sin\theta d\theta d\phi$. All of the vacua of
$\Ncal=4$ SYM on $R\times S^2$ discussed in \cite{lm} can be obtained in this
way. This limit has a clear interpretation in the T-dual picture. We start with
a configuration of a finite number of D2-branes, repeated periodically by
translating the whole arrangement by integer multiples of $2\pi k$. In the
limit $k\rightarrow\infty$, we are left with only one copy of the configuration
of D2-branes, the images being pushed off to infinity. This naturally suggests
that the dual supergravity solution is obtained by considering a configuration
of a finite number of disks. It is clear that the integers $n_i$ determine the
positions and charges of the disks in a manner analogous to the situation in
the $R\times S^3/Z_k$ theory. Again, the sizes of the disks are presumably fixed by
demanding regularity of the supergravity solutions. Note that the total sum of
the charge on the disks must equal the rank of the gauge group $N$. In the rest
of this paper, we consider non-trivial vacua of the form
$\Phi = (n,\dots,n,-n,\dots -n)$, where the integers $n$ and $-n$ each appear
$N/2$ times. In this case the dual supergravity solution is generated by the
potential $V(r,z)$ corresponding to two equal-sized disks at $z=\pm (\pi/2)n$
with charge $(\pi^2/8)(N/2)$ on each disk.

The final issue we need to discuss in this section is the normalization of the
asymptotic potential at infinity. For the SYM theory on $R\times S^2$, we can
relate $W_0$ to $g^2_{YM2}$ by using the results in \cite{lmsvv}. As discussed
in \cite{msv, lmsvv, istt} the SYM theory on $R\times S^2$ can be obtained as a
limit of the plane wave matrix model. This statement, together with the matching
of parameters in the plane wave matrix model discussed in \cite{lmsvv}, tell us
that we must have
\begin{equation}
\label{h2}
W_0 = \frac{h_2}{g^2_{YM2}},
\end{equation}
where the positive constant $h_2$ does not depend on the parameters $N$,
$g^2_{YM2}$, which define the gauge theory, and the eigenvalues of $\Phi$,
which label its vacua. For SYM theory on $R\times S^3/Z_k$, the above mentioned
relation between this theory and the theory on $R\times S^2$ suggests that we make the
identification
\begin{equation}
\label{h3}
W_0 = \frac{h_3}{g^2_{YM3}k},
\end{equation}
where $h_3$ is a positive constant that does not depend on $N, k, g^2_{YM3}$ and
the integers that label the vacua of the gauge theory.

\section{Little string theory from SYM on $R\times S^2$}
In this section, we consider in detail the supergravity solution corresponding
to the electrostatics problem for two identical disks of radius $R$ located at
$z=\pm d$ with charge $Q$ on each disk and a background potential
$W_0(r^2-2z^2)$. We wish to solve the electrostatics problem explicitly and
determine the required scaling to obtain the Lin-Maldacena NS5-brane solution.
\subsection*{The electrostatics problem for the case of two identical disks}

\label{es}
\begin{figure} \label{QD2}
\begin{picture}(0,0)%
\includegraphics{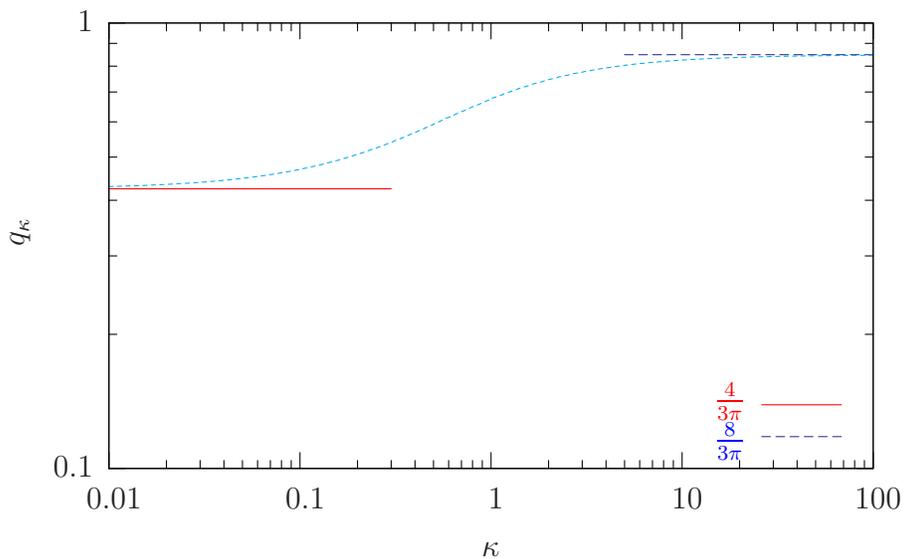}%
\end{picture}%
\begingroup
\setlength{\unitlength}{0.0200bp}%
\begin{picture}(18000,10800)(0,0)%
\put(2400,1800){\makebox(0,0)[r]{\strut{} 0.1}}%
\put(2400,10200){\makebox(0,0)[r]{\strut{} 1}}%
\put(2700,1200){\makebox(0,0){\strut{} 0.01}}%
\put(6300,1200){\makebox(0,0){\strut{} 0.1}}%
\put(9900,1200){\makebox(0,0){\strut{} 1}}%
\put(13500,1200){\makebox(0,0){\strut{} 10}}%
\put(17100,1200){\makebox(0,0){\strut{} 100}}%
\put(600,6000){\rotatebox{90}{\makebox(0,0){}\strut{$q_\kappa$}}}%
\put(9900,300){\makebox(0,0){\strut{}$\kappa$}}%
\put(14700,3075){\makebox(0,0)[r]{\textcolor{red}{\strut{}$\tfrac{4}{3\pi}$}}}%
\put(14700,2325){\makebox(0,0)[r]{\textcolor{blue}{\strut{}$\tfrac{8}{3\pi}$}}}%
\end{picture}%
\endgroup
\caption{The charge on each disk in the two-disk case. The solid and dashed
lines show the asymptotes for small and large $\kappa$ respectively.}
\end{figure}

Following the approach of
\cite{lmsvv}, we first solve the electrostatics problem for the specific case
$W_0=1$, $R=1$, $d=\kappa$ (the solution for the general case is then obtained
by linear rescaling of the coordinates and an overall rescaling of the
potential). In this case the solution must have the form
\begin{equation}
V(r,z)= (r^2 - 2z^2) + \phi_{\kappa}(r,z),
\end{equation}
where $\phi_\kappa$ is an axisymmetric solution of the Laplace
equation that vanishes at infinity. We can expand $\phi_\kappa$ in terms of
Bessel functions, and in the region between $z=-d$ and $z=d$, this expansion
takes the form
\begin{equation}
\phi_\kappa(r,z) = \int_0^\infty \frac{du}{u} e^{-u\kappa}A(u)
	\bigl(e^{-uz}+e^{uz}\bigr)J_0(ru).
\end{equation}
The potential on the two conducting disks, $\Delta$, must be constant,  and the
electric field must be continuous at all points not on the disks. Imposing
these boundary conditions leads to the following dual integral equations
\begin{equation}
\begin{split}
\int_0^\infty \frac{du}u (1+e^{-2\kappa u}) J_0(ru) A(u) &=
		\Delta-r^2 \qquad 0<r<1 \\
\int_0^\infty du J_0(ru) A(u) &= 0 \qquad r>1 .
\end{split}
\end{equation}
Following \cite{sneddon} we find that the solution of these integral equations
can be given in terms of the solution to a Fredholm integral equation of
the second kind. The problem in this case is very similar to the one considered
in \cite{lmsvv}. We have
\begin{equation}
A(u) = \frac{2u}\pi \int_0^1 dt \cos(ut) f(t),
\end{equation}
where $f(t)$ satisfies the integral equation
\begin{equation} \label{inteqn}
f(t) + \int_{-1}^1 dx K(t,x) f(x) = \Delta - 2t^2,
\end{equation}
and
\begin{equation}
K(t,x) = \frac1\pi \frac{2\kappa}{4\kappa^2+(t-x)^2}.
\end{equation}
For each value of $\Delta$, the integral equation for $f$ can be solved
numerically. From the resulting electrostatics potential, we can compute the surface charge density on the disks
\begin{equation}
\sigma(r) = \frac{1}{\pi^2}\left[\frac{f(1)}{\sqrt{1-r^2}} - \int_r^1 dt \frac{f^\prime(t)}{\sqrt{t^2-r^2}}\right].
\end{equation}
We can adjust the constant $\Delta$ until we find the value $\Delta_\kappa$ for
which the corresponding solution $f_\kappa$ satisfies $f_\kappa(1)=0$. Then the
surface charge distribution $\sigma_\kappa(r)$ for this solution vanishes at the
edge of the disks. This final condition ensures the regularity of the
corresponding supergravity solutions. The total charge on each disk is given by
\begin{equation}
q_\kappa = \frac{2}{\pi}\int_0^1 dt f_\kappa(t).
\end{equation}
Figure (\ref{QD2}) shows a plot of $q_\kappa$. For large $\kappa$ the charge on
each disk approaches $8/3\pi$, and for small $\kappa$ the charge on each disk
approaches $4/3\pi$.

Finally the solution for the general case is obtained by rescaling. The
electrostatics potential is given by
\begin{equation}
V(r,z)= W_0(r^2 - 2z^2) + W_0R^2\phi_{d/R}(r/R,z/R),
\end{equation}
and the total charge on each disk is given by
\begin{equation}
Q = W_0 R^3 q_{d/R}.
\end{equation}
\begin{figure} \label{DD2}
\begin{center}
\includegraphics{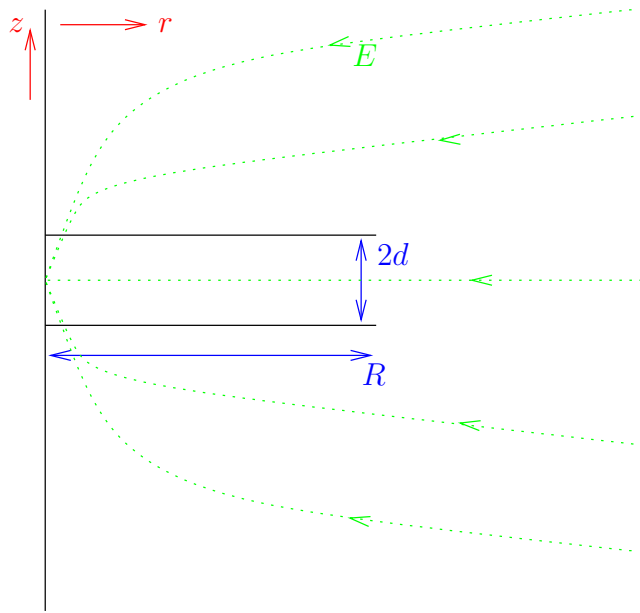}
\end{center}
\caption{The electrostatics problem for two identical disks. The dotted lines
show the background electric field configuration.}
\end{figure}
\subsection*{The limit of the Lin-Maldacena solution}
Now we can determine the limit of this solution that gives the Lin-Maldacena
solution for NS5-branes on $S^5$. In the region between the disks with
$0<r<R$, our solution is an axisymmetric solution of the Laplace
equation that is regular at $r=0$, so we can expand the solution in terms of modified Bessel
functions
\begin{equation}
V(r,z) = V_{z=d}+\sum_{n=1}^\infty c_n
	\cos \left( \tfrac{(2n+1)\pi z}{2d} \right)
	I_0 \left( \tfrac{(2n+1)\pi r}{2d} \right).
\end{equation}
The coefficients $c_n$ may be determined by using the potential at $r=R$. This
gives
\begin{equation}
c_n = \left( I_0 \left( \tfrac{(2n+1)\pi R}{2d} \right) \right)^{-1}
	2 W_0 R^2 \int_0^1 dz \cos\left(\tfrac{(2n+1)\pi z}{2}\right)
	\left( 1-2(\kappa z)^2-\Delta_\kappa+\phi_\kappa(1,\kappa z) \right).
\end{equation}
Using our numerical solution for $\phi_\kappa$, the above integral can be
performed numerically. In the limit $d\ll R$, this gives
\begin{equation}
c_1 \approx 1.56 W_0 Rd
	\left( I_0 \left( \tfrac{\pi R}{2d} \right) \right)^{-1} .
\end{equation}
For large $R/d$ this expression will be dominated by the Bessel function,
which takes the asymptotic form
\begin{equation*}
(I_0(z))^{-1} \sim \sqrt{2\pi z}e^{-z}
\end{equation*}
To preserve some non-trivial geometry, we must then scale $W_0$ exponentially.
Doing so keeps $c_1$ finite in the limit, but sends all the other coefficients
to zero so that we recover the Lin-Maldacena solution. More precisely,
the Lin-Maldacena solution is obtained in the limit
\begin{eqnarray}
	R\rightarrow \infty \hspace{2cm} d\ \mathrm{fixed} \hspace{1.5cm} & W_0 \sim R^{-1}(Rd)^{-1/2} e^{\tfrac{\pi R}{2d}}.
\end{eqnarray}

\subsection*{The gauge theory interpretation}
Having understood the correct scaling on the gravity side, we can translate
this into a condition on the gauge theory parameters. This amounts to
\begin{equation}
N\to\infty \qquad n \; \mathrm{fixed} \qquad
	 \frac{1}{g_{YM2}^2}\lambda^{1/2}n^{1/2}e^{-b\lambda^{1/3}/n}
		\;\mathrm{fixed},
\end{equation}
where the 't Hooft coupling is $\lambda = g^2_{YM2}N$ and $b$ is a numerical coefficient related to the constant appearing in (\ref{h2}) by $b=(\pi/4)(3/h_2)^{1/2}$. We see that this is a large $N$
limit, where the 't Hooft coupling is also scaled to infinity in a controlled
way, and is very similar to the limit that was found in the case of the PWMM
in \cite{lmsvv}.  Note that the number of NS5-branes is $N_5 = 2n$.  

\section{Little string theory from $\mathcal{N}=4$ SYM on $R\times S^3/Z_k$}
Now we wish to perform a similar detailed analysis for the supergravity
solution corresponding to a periodic array of disks of radius $R$, where the
disks are located at $z=(2m+1)d$ ($m$ is any integer), the charge on each disk
is $Q$, and the background electric field is given by the potential
$W_0(r^2-2z^2)$. Again we first solve the electrostatics problem, then find the
limit that recovers the NS5-brane solution of Lin and Maldacena.

\subsection*{The electrostatics problem for a periodic array of disks}
\begin{figure} \label{SSYM}
\begin{center}
\includegraphics{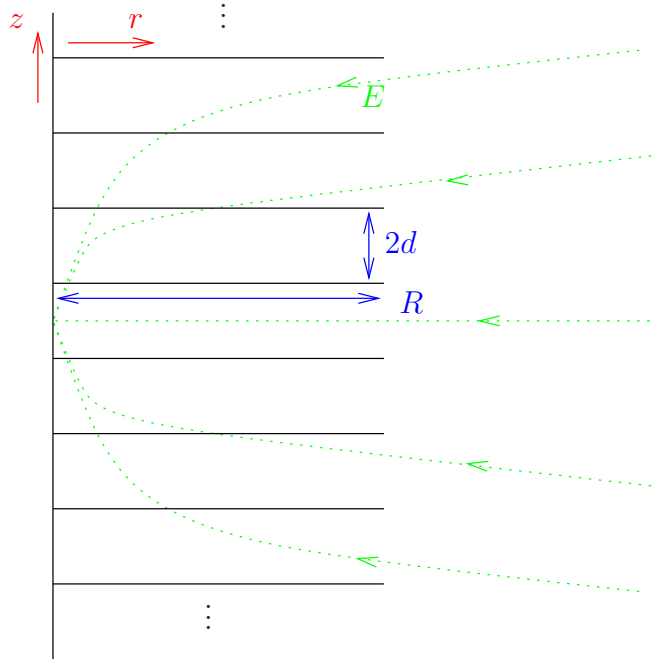}
\end{center}
\caption{The electrostatics problem in the case a periodic array of disks.
The dotted lines show the background electric field configuration.}
\end{figure}

As in the previous section, we solve the electrostatics problem for the special
case $W_0=1$, $R=1$, $d=\kappa$, and obtain the solution for the general case by
rescaling. In the absence of the background potential the charge distribution
on each disk will be the same. Adding the background field will affect the
charge distribution on each disk, but since the radial part of the electric
field it creates is identical on each disk, the charge distribution will remain 
the same on each disk (see figure \ref{SSYM}).

We can separate the potential into the sum of the background field and the part
due to the charge on the disks.
\begin{equation}
V = r^2-2z^2 + \phi_\kappa(r,z),
\end{equation}
where $\phi_\kappa$ is periodic in $z$ because the charge on each disk is
identical. Formally, we can expand $\phi_\kappa(r,z)$ in terms of Bessel
functions as
\begin{equation}
\label{phiexp}
\phi_\kappa(r,z) = \int_0^\infty \frac{du}{u} J_0(ru) A(u)
	\sum_{n=-\infty}^{\infty} e^{-u|(2n+1)\kappa-z|},
\end{equation}
and then try to determine the function $A(u)$ by the imposing the boundary
conditions. If we take the value of the potential $V$ to be $\Delta-2\kappa^2$
on the disk at $z=\kappa$, then by imposing the boundary conditions
we obtain the following dual integral equations
\begin{equation}
\begin{split}
\int_0^\infty \frac{du}u \left(1+\frac{2e^{-2\kappa u}}{1-e^{-2\kappa u}}\right)
	J_0(ru) A(u) &= \Delta-r^2 \qquad 0<r<1 \label{firstdual}\\
\int_0^\infty du J_0(ru) A(u) &= 0 \qquad r>1.
\end{split}
\end{equation}
However, direct attempts to solve these equations are met with divergences and
various difficulties. The reason is that these equations hold only formally,
because the sum in the expression for the potential \eqref{phiexp} actually
diverges. Physically, there is no divergence because the electric field remains
finite. This is the same type of situation encountered for an infinite number of
equally spaced point charges (or an infinite line of charge) on the $z$-axis,
which occurs simply because we try to express the potential as a sum of the
Coulomb potential for each charge. If we consider the potential difference
between any two points, there is no divergence, so we can regularize
\eqref{phiexp} by subtracting the potential at any fixed reference point. 

\begin{figure} \label{QSYM}
\begin{picture}(0,0)%
\includegraphics{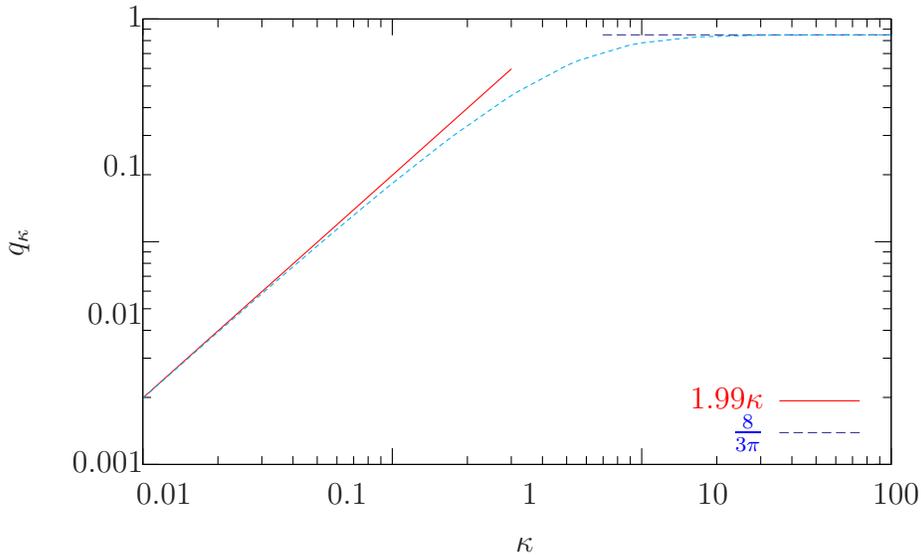}%
\end{picture}%
\begingroup
\setlength{\unitlength}{0.0200bp}%
\begin{picture}(18000,10800)(0,0)%
\put(3000,1800){\makebox(0,0)[r]{\strut{} 0.001}}%
\put(3000,4600){\makebox(0,0)[r]{\strut{} 0.01}}%
\put(3000,7400){\makebox(0,0)[r]{\strut{} 0.1}}%
\put(3000,10200){\makebox(0,0)[r]{\strut{} 1}}%
\put(3300,1200){\makebox(0,0){\strut{} 0.01}}%
\put(6750,1200){\makebox(0,0){\strut{} 0.1}}%
\put(10200,1200){\makebox(0,0){\strut{} 1}}%
\put(13650,1200){\makebox(0,0){\strut{} 10}}%
\put(17100,1200){\makebox(0,0){\strut{} 100}}%
\put(600,6000){\rotatebox{90}{\makebox(0,0){\strut{}$q_\kappa$}}}%
\put(10200,300){\makebox(0,0){\strut{}$\kappa$}}%
\put(14700,3000){\makebox(0,0)[r]{\strut{}\textcolor{red}{$1.99\kappa$}}}%
\put(14700,2400){\makebox(0,0)[r]{\strut{}\textcolor{blue}{$\tfrac{8}{3\pi}$}}}%
\end{picture}%
\endgroup
\caption{The charge on a disk as a function of the spacing between disks. The
numerical result is given by the dashed line. The solid line is the asymptotic
behaviour for small $\kappa$, $q\sim 1.99\kappa$. For large $\kappa$ the charge
approaches $\tfrac{8}{3\pi}$.}
\end{figure}

In this case, it is more convenient to consider the first integral equation
\eqref{firstdual} as a condition on the electric field rather than the electric
potential
\begin{equation}
\int_0^\infty du \left(1+\frac{2e^{-2\kappa u}}{1-e^{-2\kappa u}}\right)
        J_1(ru) A(u) = 2 r \qquad 0<r<1.
\end{equation}
The dual integral equations can then be solved by introducing a function
satisfying a Fredholm integral equation of the section kind,
\begin{equation} \label{inteq2}
f_\kappa(x) + \int_0^1 du K(x,u) f_\kappa(u) = -\frac{8x}{\sqrt\pi},
\end{equation}
where
\begin{equation}
\label{persoln}
A(u) = -\frac{1}{\sqrt\pi} \int_0^1 d\xi \, \sin(u \xi) f_\kappa(\xi).
\end{equation}
The kernel is given by
\begin{equation}
K(x,u) = \frac 1\pi \int_0^\infty dt k(t)(-\cos(u+x)t+\cos|u-x|t),
\end{equation}
where
\begin{equation}
k(u) = \frac{2e^{-2\kappa u}}{1-e^{-2\kappa u}}.
\end{equation}
These integrals can be evaluated and the result is
\begin{equation}
\begin{split}
K(x,u)=\frac{1}{2\pi\kappa}
        \Bigl(&
        \Psi\Bigl(1+\frac{i(x+u)}{2\kappa}\Bigr)
        +\Psi\Bigl(1-\frac{i(x+u)}{2\kappa}\Bigr) \\
        &-\Psi\Bigl(1+\frac{i|x-u|}{2\kappa}\Bigr)
        -\Psi\Bigl(1-\frac{i|x-u|}{2\kappa}\Bigr)
        \Bigr),
\end{split}
\end{equation}
where $\Psi$ is the digamma function. We solved \eqref{inteq2} numerically
using the Nystr\"om method (e.g.~\cite{numbook}). In contrast to the two disk
case, since we considered the integral equation corresponding to a condition on
the electric field, there is no $\Delta$ to adjust to ensure that the surface
charge density at the edge of the disk vanishes. In fact, for the form of the
solution given in (\ref{persoln}), this condition is automatically satisfied as
long as $f_\kappa$ is bounded. In terms of $f_\kappa$, the charge on each disk
is
\begin{equation}
q_\kappa = -\frac{1}{\sqrt{\kappa}}\int_0^1 dt\ t f_\kappa(t).
\end{equation}
Using our numerical solution for $f_\kappa$ we found that $q_\kappa$ approaches
$8/3\pi$ for large $\kappa$ and approximately $1.99\kappa$ for small $\kappa$
(see figure \ref{QSYM}).

In principle, it is possible to determine the regularized potential completely
from this solution for $f_\kappa$ (however, the integrals involved are rather
computationally expensive). Then the potential for the case of general $W_0$,
$R$ and $d$ is obtained by a linear rescaling of coordinates and an overall
rescaling of the potential. Specifically, we note that the charge on each disk
in the general case is
\begin{equation}
Q= W_0R^3 q_{d/R}.
\end{equation}

\subsection*{The limit of the Lin-Maldacena solution}
To determine how the Fourier coefficients of the potential scale with $\kappa$,
we found it was most efficient to use the method of conformal mapping. Near the
edge of the disks, when their radial size is much larger than their separation,
the electrostatics problem becomes two-dimensional. By defining the complex
coordinates $\zeta=(r-R)+iz$, and $w=2\partial_\zeta V$ any holomorphic
function $w(\zeta)$ will be a solution of the Laplace equation. As described in
\cite{lm} the appropriate mapping in this case is
\begin{equation}
\partial_w \zeta = \alpha \tanh\bigl(\frac{\pi w}{\beta} \bigr)
\end{equation}
and so
\begin{equation}
\zeta=\frac{\alpha\beta}\pi\log\cosh\left(\frac{\pi w}{\beta}\right),
\end{equation}
where $\alpha$, $\beta$ are constants. Inverting this we find
\begin{equation}
w=\frac{\beta}\pi\cosh^{-1}\left(e^{\frac{\pi \zeta}{\alpha \beta}}\right).
\end{equation}
If we fix the positions of the disks to be at $\zeta=i(2md)$, where $m$ is an
integer, we have $d=\alpha \beta/2$. The vertical electric field at any disk
should be $-4W_0\Im(\zeta)$, so that $\beta=8W_0 d$ and $\alpha=1/4W_0$.

Expanding the potential in terms of modified Bessel functions, as in the
two-disk case, we find that
\begin{equation}
c_1 \approx \frac{16W_0d^2}{\pi}(I_0(\tfrac{\pi R}{2d}))^{-1}(0.659).
\end{equation}
Again, therefore, to preserve non-trivial geometry we must scale $W_0$
exponentially. The precise scaling form to obtain the Lin-Maldacena solution is
\begin{eqnarray}
	R\rightarrow \infty \hspace{2cm} d\ \mathrm{fixed} \hspace{1.5cm} & W_0 \sim R^{-1/2}d^{-3/2} e^{\tfrac{\pi R}{2d}}.
\end{eqnarray}
\subsection*{The gauge theory interpretation}
In terms of the gauge theory parameters, we have
\begin{equation}
N\to\infty \qquad k \; \mathrm{fixed} \qquad
	 \frac{1}{g_{YM3}^2}\lambda^{1/4}k^{1/2}e^{-c\lambda^{1/2}/k}\;
		\mathrm{fixed},
\end{equation}
where the 't Hooft coupling is $\lambda = g^2_{YM3}N$ and $c$ is a numerical coefficient related to the constant appearing in (\ref{h3}) by $c=(2\pi/1.99 h_3)^{1/2}$.  This is again a double-scaling limit
in which the 't Hooft coupling is scaled to infinity in a controlled way.  Note
that the number of NS5-branes in this case is $N_5=k$.

\section{Discussion}
We have given an explicit prescription for taking double-scaling limits of SYM
theory on $R\times S^2$ and $\Ncal=4$ SYM on $R\times S^3/Z_k$ to obtain little
string theory on $S^5$. These limits were obtained by using the family of
supergravity solutions found by Lin and Maldacena \cite{lm}. With the similar
result in \cite{lmsvv}, we have demonstrated that it is possible to take such a
limit in each of the three generic examples of this family of solutions, and
in each of the three field theories to which they are dual.

In each case, the precise form of the double-scaling limit is similar. Whereas
in the plane wave matrix model it was found the correct limit was \cite{lmsvv}
\begin{equation}
N_2\to\infty \qquad N_5 \; \mathrm{fixed} \qquad
	N_2 \sim \lambda^{5/8} e^{a\lambda^{1/4}/N_5} \;,
\end{equation}
we found above that for the SYM theory on $R\times S^2$ we have
\begin{equation}
N\to\infty \qquad n \; \mathrm{fixed} \qquad
	 N \sim \lambda^{1/2}n^{-1/2}e^{b\lambda^{1/3}/n}\;,
\end{equation}
and for $\Ncal=4$ SYM theory on $R\times S^3/Z_k$ we have
\begin{equation}
N\to\infty \qquad k \; \mathrm{fixed} \qquad
	N\sim \lambda^{3/4}k^{-1/2}e^{c\lambda^{1/2}/k}\;.
\end{equation}

As noted in \cite{lmsvv}, it is sensible that the correct limit to obtain
little string theory from these field theories is a double-scaling limit as
opposed to a strict 't Hooft limit. If the correct limit was the 't Hooft limit,
then it would seem strange that the field theory could produce string loop
interactions. That the 't Hooft coupling should also be scaled to infinity in a
controlled way allows the field theory to reproduce the string genus expansion.

Suppose we consider the genus expansion for some physical observable in one
of these theories
\begin{equation}
F = \sum_g N^{2-2g} f_g(\lambda,\alpha) ,
\end{equation}
where $\alpha$ represents the other parameters. The double-scaling limit should
be such that all terms in this expansion contribute. For this to occur, the
terms in the expansion would have to take a particular form when $\lambda$ is
large. In the case of the PWMM, this form was found to be \cite{lmsvv}
\begin{equation}
f_g(\lambda) \to a_g \bigl( \lambda^{5/8} e^{a\lambda^{1/4}/N_5} \bigr)^{2g-2},
\end{equation}
where the bracketed expression divided by $N_2$ serves as the effective
coupling constant. Here we find for the SYM theory on $R\times S^2$ we must have
\begin{equation}
f_g(\lambda) \to a_g \bigl( \lambda^{1/2} e^{b\lambda^{1/3}/n} \bigr)^{2g-2},
\end{equation}
and for $\Ncal=4$ SYM theory on $R\times S^3/Z_k$
\begin{equation}
f_g(\lambda) \to a_g \bigl( \lambda^{3/4} e^{c\lambda^{1/2}/k} \bigr)^{2g-2}.
\end{equation}
Interestingly, although these field theories live in different numbers of
dimensions, it is possible to recover little string theory from each of them
by similar double-scaling limits.

Obvious difficulties arise in checking these predictions. One might hope that
there are some BPS observables for which such a check might be feasible.
In the case of the circular Wilson loop in $\Ncal=4$ SYM the full set of planar
diagrams can be summed \cite{esz}. The result in that case took the form
\begin{equation}
\bigl< W\bigr>_{N=\infty}=\sqrt{\frac{2}{\pi}}\lambda^{-3/4}e^{\sqrt\lambda}.
\end{equation}
This result has been extended to all orders in \cite{dg}, where it was shown
that the asymptotic behaviour goes like $e^{\sqrt{\lambda}}$ at each order. That
behaviour also arises from modified Bessel functions. It would be interesting
to calculate the circular Wilson loop in $\Ncal=4$ SYM on $R\times S^3/Z_k$, and
to compare it with our results here.

Other open questions remain. For example, as noted in \cite{lmsvv}, the solution
for little string theory on $S^5$ given by Lin and Maldacena \cite{lm} is the
simplest of an infinite family of solutions that have an infinite throat with
$H$-flux. It would be interesting to understand if these solutions could arise
from limits of more general disk configurations.  It would also be
interesting to understand more about the vacua of little string theory dual to
these solutions.

\section*{Acknowledgements}
We would like to thank Donovan Young for pointing out reference \cite{dg}, Hai
Lin for helpful comments and especially Mark Van Raamsdonk for many helpful
discussions. This work has been supported in part by the Natural Sciences and
Engineering Council of Canada and the Killam Trusts.

\end{document}